\documentclass{aa}
\usepackage{epsfig,amsmath,amssymb}

\def\msol{M_\odot}

\def\te{T_{\rm eff}}

\def\taud{\tau_{\rm growth}}
\def\tdeut{\tau_{\rm D}}
\def\enuc{\epsilon_{\rm nuc}}
\def\simgr{\,\hbox{\hbox{$ > $}\kern -0.8em \lower 1.0ex\hbox{$\sim$}}\,}
\def\simle{\,\hbox{\hbox{$ < $}\kern -0.8em \lower 1.0ex\hbox{$\sim$}}\,}
\def\beq{\begin{equation}}
\def\eeq{\end{equation}}

\begin{document}


\title{Pulsating young brown dwarfs}
 \author{Francesco Palla\inst{1} and Isabelle Baraffe\inst{2,3}
}


\institute{INAF-Osservatorio Astrofisico di Arcetri, Largo E. Fermi 5, 50125
Firenze, Italy (palla@arcetri.astro.it)
\and
C.R.A.L (UMR 5574 CNRS),
 Ecole Normale Sup\'erieure, 69364 Lyon
Cedex 07, France (ibaraffe@ens-lyon.fr) \and
Astrophysikalisches Institut Potsdam, An der Sternwarte 16,
14482 Potsdam, Brandenburg, Germany
}

\date{Received /Accepted}

\titlerunning{}
\authorrunning{Palla and Baraffe}
\abstract{We present the results of a nonadiabatic, linear stability analysis
of models of very low-mass stars (VLMSs) and brown dwarfs (BDs) during the
deuterium burning phase in the center. We find unstable fundamental modes
with periods varying between $\sim$5 hr for a 0.1 $\msol$ star and $\sim$1 hr
for a 0.02 $\msol$ BD. The growth time of the instability decreases with
decreasing mass and remains well below the deuterium burning time scale 
in the mass range considered (0.1--0.02 M$_\odot$).
These results are robust against variations of the
relevant input physics in the evolutionary models. We identify 
possible candidates for pulsational variability among known VLMSs and BDs in
nearby star forming regions whose location in the HR diagram falls within
or close to the boundary of the instability strip. Finally, we discuss the
possibility that the variability observed in a few objects with periods of 
$\sim$1~hr can be interpreted in terms of pulsation.
}

\maketitle

\section{Introduction}

Stellar pulsation represents a unique probe to study the internal structure
and evolution of stars. This technique has been applied to many classes of
stars, both on the main sequence and in earlier/later stages of their
evolution (e.g., Gautschy \& Saio 1995, Marconi \& Palla 1998).  New classes
of pulsators are being discovered across the HR diagram from ground-based
observations and with the advent of space missions (MONS, COROT)
the potential of discovering solar-like oscillations in other stars promises
a major improvement in our understanding of stellar interiors (e.g.,
Thompson, Cunha \& Monteiro 2003).

In this Letter, we propose the identification of a new class of pulsators,
VLMSs and BDs, in which the instability is induced by deuterium (D-) burning
in their centers. Due to the high sensitivity of nuclear reactions on
temperature, fully convective stars can become pulsationally unstable during
the burning phases.  Gabriel (1964) was the first to suggest that such
mechanism, called $\epsilon$-mechanism, would make low-mass stars on the main
sequence vibrationally unstable. 
Later on, Toma (1972) showed that
pre--main-sequence (PMS) stars in the range 0.2--2.0 $\msol$ could also
become pulsationally unstable during the D-burning phase and suggested a
possible relation of this phenomenon with the observed variability of T Tauri
stars.  In both cases, however, and for different reasons, the suggestion has
not met with success and to date there is no observational evidence for the
$\epsilon$-mechanism in any class of pulsating star where the destabilization
is mainly induced by opacity effects in ionization regions near the
stellar surface (the $\kappa$-mechanism).
\\ \indent
Indeed, no periodic variability, which could be interpreted as 
pulsation, has been reported up to now in main sequence low-mass stars,
which may  not be surprising, given 
the weak dependence of the $pp$-chains on temperature and the resulting long
$e$-folding times for the $\epsilon$-mechanism.
For young PMS stars, the main difficulty to confirm observationally
the results of Toma (1972) is due to the very short D-burning time scale,
$\simle \, 1 $ Myr, for stars with mass $\ge$ 0.2 $\msol$. Moreover,
the results obtained by Toma (1972) are optimised, in terms
of D-burning time scale and e-folding time for the pulsation, since 
he adopted an initial D-abundance
about ten times higher than the currently measured interstellar
value and ignored
D-depletion during protostellar accretion that can substantially reduce its
abundance at the beginning of the PMS phase of low-mass stars. 
As we will show below, VLMSs ($\simle$0.1~$\msol$) and BDs in the
earliest evolutionary stages have the appropriate physical conditions to
circumvent the limitations of previous attempts.

\section{Instability induced by D-burning in VLMSs and BDs}

BDs are expected to begin the PMS phase with convective interiors and with
the full amount of interstellar deuterium since during the preceding phase of
protostellar accretion their centers are too cold to start any nuclear
reaction. The ignition of deuterium requires temperatures around $10^6$~K
that can only be reached in protostars more massive than $\sim$0.1--0.2
$\msol$, depending on the mass accretion rate (Stahler 1988).  Thus, very
low-mass objects need to contract somewhat in the PMS phase before deuterium
can start burning and being depleted. Once the critical temperature is
achieved, the D-burning phase occurs on a time scale that varies between
$\tau_D\sim$2.5~Myr for a 0.1~$\msol$ star and
$\tau_D\sim$20~Myr for a 0.02~$\msol$ brown dwarf (Chabrier et al. 2000).

Since the energy generation rate, $\epsilon_{\rm nuc}$, scales approximately
with the 12-th power of the temperature, the rate of combustion is slower for the
lowest mass objects.  The same temperature sensitivity of $\epsilon_{\rm nuc}$
is at the root of the instability induced by any temperature variation: since
$\delta \epsilon_{\rm nuc}/\epsilon_{\rm nuc} \sim 12 \,\delta T/T$, a small
T-perturbation induces a variation of $\epsilon_{\rm nuc}$ which is an order
of magnitude bigger. (Note that $\epsilon_{\rm nuc}$ is also linearly
proportional to density, but no destabilizing effect is expected from
variations in this quantity).  In terms of pulsation analysis, the time scale
for the growth of the instability, $\tau_{\rm growth}$, is inversely
proportional to $\delta \epsilon_{\rm nuc}$ and should be shorter than the
D-burning time for the $\epsilon$-mechanism to operate.  Therefore, in order
to test its viability, it is important to follow numerically the growth of
the instability in time since the onset of the D-burning phase.

\section{Evolutionary models  and pulsation analysis}

Our analysis focuses on objects in the mass range 0.1--0.02 $\msol$, close to
the minimum value for D-burning (0.013 $\msol$, Chabrier et al. 2000).
The evolutionary models are based on the input physics described in Baraffe et
al. (1998) and on the so-called ``NextGen" grainless
atmosphere models (Hauschildt et al. 1999).  Models are calculated for a
solar metallicity, with an initial deuterium mass fraction $D_0 \, = \,
2\times 10^{-5}$, characteristic of the local interstellar medium (Linsky
1998), and with mixing length parameter equal to the pressure scale height.
During the early phase of D-burning, brown dwarfs are relatively hot, with
$\te \,  \simgr \, 2300K$, and their atmospheric properties are still well
described by dust-free models.

\subsection{Linear stability analysis results}

We performed a non-adiabatic linear stability analysis of the model
structures during the D-burning phase, searching for the presence of unstable
radial eigen modes.  Unstable modes characterize oscillations around the
hydrostatic equilibrium configuration. If perturbations have time to grow,
they could reach large amplitudes and result in  periodic phases of expansion
and contraction, with a pulsation period  related to the dynamical time scale
of the object $\tau_{\rm dyn} \sim (G \bar \rho)^{-1/2}$.  References on
the pulsation code used for the present purpose can be found in Lee (1985)
and Baraffe et al. (2001).  In all models analyzed, we find unstable
fundamental modes, the results being rather robust against variations of
the relevant input physics (see Sect. 3.2).  We also find
unstable first overtones, but their stability properties are sensitive
to the input physics and their growth rate is usually smaller than that
of the fundamental mode.  Higher overtones remain stable during the whole
D-burning phase. We thus restrict the rest of our study to fundamental modes.

\begin{figure}
\psfig{file=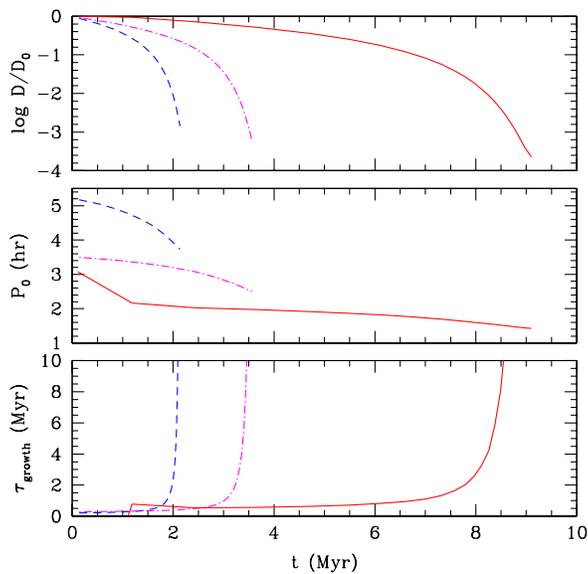,height=80mm,width=80mm} 
\caption{ Evolution of the D-abundance (in units of the initial
abundance $D_0=2\times 10^{-5}$ in mass fraction), the fundamental mode
period $P_0$ (in hr) and growth time scale $\taud$ (in Myr) as a function of
time during the D-burning phase of a 0.1 $\msol$ star (dashed lines),
and a 0.06 $\msol$ (dash-dotted line) and 0.03 $\msol$ (solid line) BD. Note
that only unstable models are shown. 
} 
\label{fig_tl} 
\end{figure}

\begin{figure}
\psfig{file=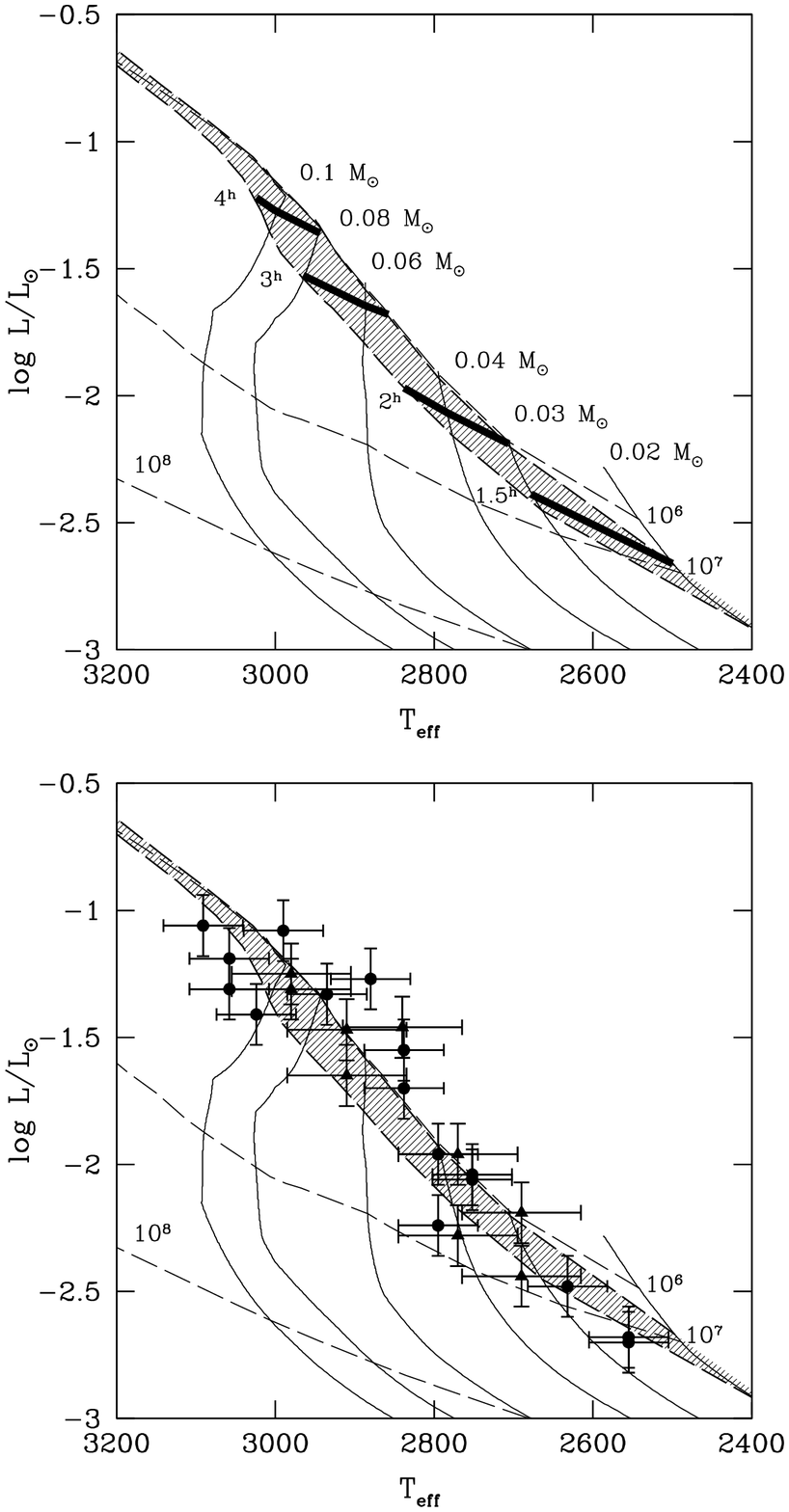,height=138mm,width=88mm} 
\caption{{\it Top panel:} Location of the D-instability strip (shaded
area) in the HR diagram.
Tracks for different masses and isochrones are indicated as 
labeled. The isoperiod curves (in hr) are shown by the thick lines
within the strip. {\it Bottom panel:} Distribution of known VLMSs and
BDs in Taurus (circles) and Cha I (triangles). Data points for Taurus are
from Brice\~no et al. (2002), Luhman et al. (2003), and Luhman (2004), and
for Cha I from Comer\'on et al. (2000).}

\label{fig_hrd} 
\end{figure}

In the mass range of interest, fundamental mode
periods $P_0$ vary between  $\sim$1hr and $\sim$5 hr, close to $\tau_{\rm
dyn}$.  For a given mass, $P_0$ decreases during the D-burning phase, since
contraction proceeds and $\bar \rho$ increases.  This is illustrated in Table
1 which summarizes the characteristics of VLMSs and BDs and their main
pulsation properties during the D-burning phase for different masses. We
define $\taud$ as the e-folding time characterizing the growth time scale of
the pulsation amplitude for the fundamental mode.  Fig.~\ref{fig_tl}
displays $P_0$ and $\taud$ for a 0.1 $\msol$ VLMS, and 0.06 and 0.03
$\msol$ BDs.  Figure~\ref{fig_tl} and Table 1 show that at the beginning of
the D-burning phase, $\taud$ is shorter, by at least a factor 10, than the
D-burning time scale $\tdeut$, which is 2.5, 4.2  and 9.3 Myr
for 0.1, 0.06 and 0.03 $\msol$, respectively. Note
that only unstable models are shown in Fig. \ref{fig_tl}, since the
structure becomes
stable again before complete depletion of D.
It remains small during a significant fraction
of the D-burning phase in the lowest mass BDs, indicating that the
perturbations may have time to grow to reach significant amplitudes.  The
location of the D-instability strip in the HR diagram is shown in the top
panel of Fig.~\ref{fig_hrd}, along with evolutionary tracks of selected
masses and isochrones. The curves of constant period cut the strip almost
horizontally (nearly constant luminosity) and the shortest periods are found
for the least massive BDs. Note that our models can also provide the locus of
the D-instability strip in the color-magnitude diagram (CMD) for different
bands for comparison with observations. We plan to present 
such diagrams in a forthcoming paper.

 \begin{figure}
\psfig{file=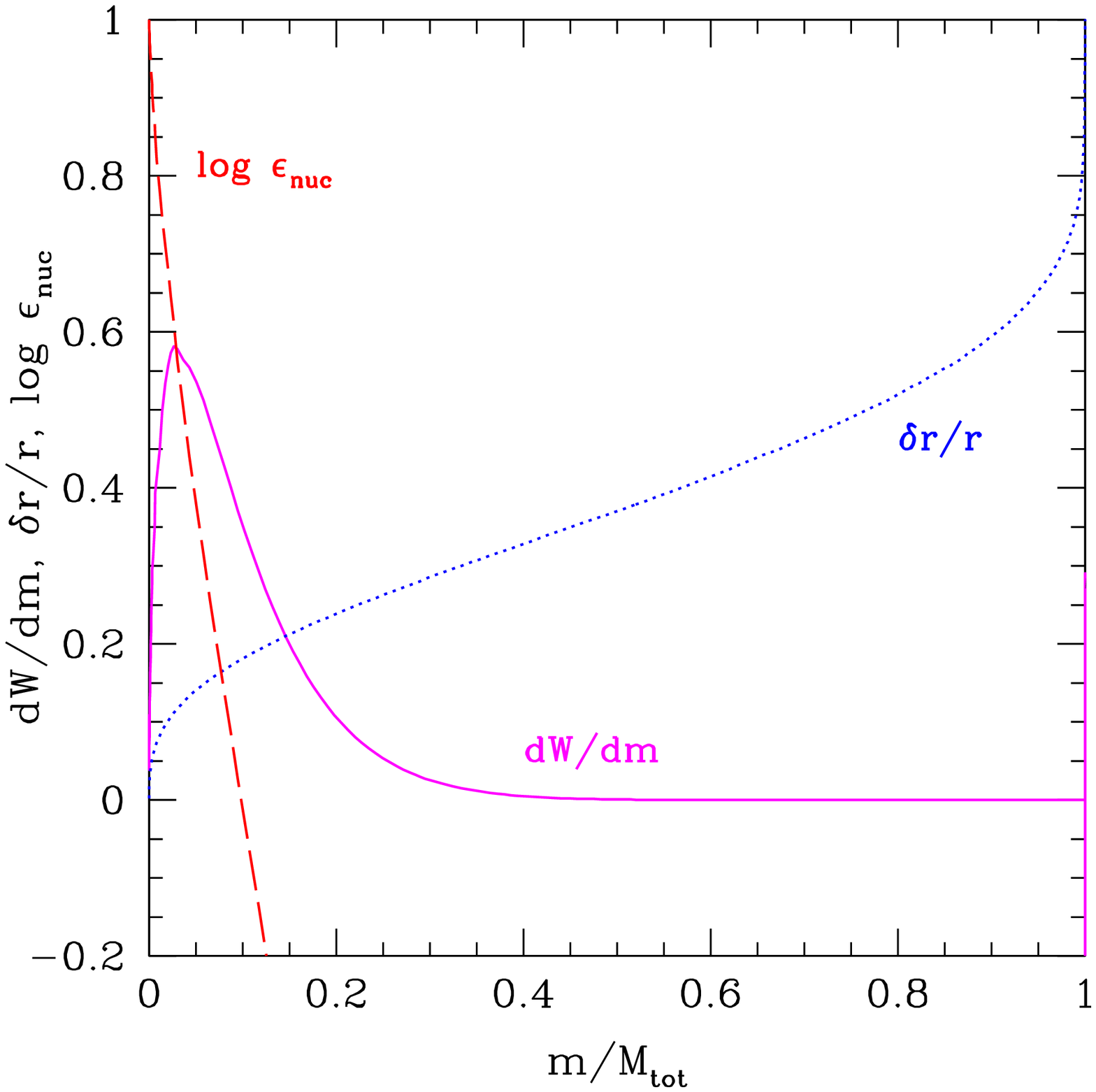,height=80mm,width=80mm} 
\caption{Differential work integral $dW/dm$, in arbitrary units (solid line),
as a function of interior mass (in units of the total mass) for a
0.03~$\msol$ BD at the beginning of the D-burning phase (D-abundance
$D/D_0$=0.71, age=2.4 Myr). The pulsation amplitude $\delta r/r$ (dotted
line), arbitrarily normalized to 1 at the surface, and the nuclear energy
generation rate $\enuc$ (in erg/g/s, dashed line) are also displayed.  }
\label{fig_tw} \end{figure}

Inspection of the differential work integral  dW/dm throughout the structure
of D-burning objects
 reveals the exciting (positive dW/dm) and the damping (negative dW/dm)
zones. This quantity is plotted in Fig. \ref{fig_tw} for the 0.03 $\msol$ BD
at the beginning of the D-burning phase. The positive work in the central
region is due to the perturbation of the nuclear energy generation rate and
characterizes the $\epsilon$-mechanism. As a rough approximation, the work
integral in the central region follows

\beq
W \propto \enuc  \epsilon_T, 
\eeq
\noindent
with $\epsilon_{\rm nuc}$ the nuclear energy generation rate 
and $\epsilon_T \equiv ({\partial ln \, \enuc \over \partial ln
\,T})_\rho$.  The source of nuclear energy is due
to the $^2H(p,\gamma)^3He$ reaction with $\epsilon_T \, \approx \, 12-14$ in
our objects, at temperatures $T \, \approx \, 10^6$ K.  It is the combination
of such high sensitivity of the nuclear energy generation rate with
temperature, and the fully convective structure of low-mass objects, 
which yields a less centrally condensed structure, that
allows the excitation by the $\epsilon$-mechanism. Indeed, VLMSs and BDs are
characterized by a low central condensation with a ratio of central to mean
density $\rho_{\rm c}/\bar\rho  \approx  4.5-6$. This ratio decreases with
mass and for a given mass it remains nearly constant with age.  The
$\epsilon$-mechanism is usually unimportant for radial pulsation in stars
because of the small amplitudes that can develop in  dense central regions
where nuclear burning occurs.  However, stars less centrally condensed are 
thought to be  potential sites for the $\epsilon$-mechanism (Gabriel
1964).  Finite amplitudes of the eigenfunctions can indeed be
expected in the central regions, as illustrated by the behavior of the
pulsation amplitude $\delta r/r$ displayed in Fig. \ref{fig_tw}.

\subsection{Uncertainties of the results}

In order to analyze the sensitivity of the present instability to input
physics, we have computed evolutionary models using atmosphere models based
on different molecular linelists for TiO and H$_2$O  (see Chabrier et al.
2000 for more details).  The main effect is a change of the effective
temperature (by $\sim$100 K), and of the D-burning time scale. As a test on
the initial conditions, we have constructed evolutionary sequences starting
at temperatures $T_{\rm c}<5 \times 10^5$ K before central D-ignition.
Moreover, we have explored the effect of the mixing length ($L_{\rm
mix}=1-2\, H_{\rm P}$), since it can affect the early evolution of VLMSs and
BDs (see Baraffe et al. 2002).  All these variations yield only small
quantitative differences, mainly in the D-burning lifetime and the pulsation
growth time scale. However, the instability of the fundamental mode survives
in all cases, with $\taud \, \ll \, \tdeut$ during most of the D-burning
phase.  Finally, adopting for the $^2H(p,\gamma)^3He$ reaction the rate of
Caughlan \& Fowler (1988) or that of the NACRE compilation (Angulo et al.
1999) has no consequence.

An important uncertainty stems from the treatment of convection, which is
assumed frozen in the linear stability analysis, having neglected the
perturbation of the convective flux in the linearized energy equation.
Such standard approximation is only justified by the argument that the
convective time scale in the inner structure of our objects ($\tau_{\rm conv}
\approx 10^6-10^8$ s ) is significantly larger than the pulsation periods
($\simle 10^4$ s).  Analysis of the effect of convection on the perturbation
is beyond the scope of this study and in any case can only be qualitative
given our poor knowledge of the interaction between convection and
pulsation.

\section{Discussion}

Our results show that the perturbations excited in the center of D-burning
VLMSs and BDs have smaller growth time scale than that of the D-burning
phase and may thus have time to grow to the level of observable amplitudes.
Unfortunately, our linear models cannot be used to estimate the amplitude of
the pulsations.  More theoretical work is required and we hope that these
initial results will stimulate the development of non-linear calculations in
order to estimate pulsation amplitudes and of fully hydrodynamical models to
construct synthetic light curves for comparison with observations.  In any
case, it is important to realize that it may be difficult to distinguish
pulsations from other sources of photometric variability, such as rotation
and spots or atmospheric events, that occur in VLMSs and BDs on similar time
scales.  
Fortunately,
numerous candidates exist which could provide good targets for detailed
observational studies.  

As an illustration, the lower panel of Fig.~\ref{fig_hrd} displays the
location in the HR diagram of 25 spectroscopically confirmed VLMSs and BDs
drawn from larger surveys in Taurus and Chamaeleon I (Brice\~no et al. 2002,
Comer\'on et al. 2000, Luhman et al.  2003, Luhman 2004).  The rather large
error bars indicate the current uncertainties in the conversion from spectral
types to effective temperatures (see discussion in, e.g., Luhman 2004), but
the overall distribution closely matches the predicted position of the
instability strip down to the lowest masses with the shortest pulsational
periods.

Evidence for photometric variability of young VLMSs and BDs
both in the field and in young clusters has been already obtained by several
groups in the last few years (Bailer-Jones \& Mundt 2001, Joergens et
al. 2003, Zapatero Osorio et al. 2003). The amplitude of the observed
variations vary between tens of mmag in the optical to $\sim$0.05--0.2 mag in
the near-IR where these cool objects emit most of their energy.  In several
cases,  periodic variability has been reported with tentative periods in the
range from half an hour to several hours. Very interesting results have been
found in the $\sigma$ Orionis clusters (age $\sim$3 Myr) where two objects (S
Ori 27 and S Ori 28) have periods around 3~hr and even shorter in the case of
S~Ori~31 ($\sim$2 hr) and S~Ori~45 ($\sim$0.75 hr) (Bailer-Jones \&
Mundt 2001, Zapatero-Osorio et al. 2003).
The minimum fundamental periods that we find are close
to $\sim$~1hr, corresponding to the lowest mass studied (cf. Table 1).
Although the period of  S Ori 45 is slightly shorter than 1 hr, 
it is compelling that this object has precisely a mass
which could be as low as 0.02 $\msol$ (B\'ejar et al. 1999).
Several other candidates in the $\epsilon$ Orionis cluster appear to share
similar properties (Scholz \& Eisl\"offel 2004). So far, the observed 
variability has been interpreted in terms of rotation periods, presence
of photospheric cool or hot spots, interaction with accretion disks, and/or
atmospheric events. For objects with the shortest periods 
($\sim$1~hr), the inferred
rotational velocities would exceed $\sim$100 km~s$^{-1}$ at or above
breakup (e.g., Joergens et al. 2003).
The pulsational instability induced by D-burning thus offers
a new interpretation that relies on fundamental (sub)stellar properties. 
It also has the potential of providing direct information on the 
otherwise inaccesible internal structure of VLMSs and BDs.
More fundamentally, it could
provide the first evidence for the existence of the $\epsilon$-mechanism.

\begin{table*}
\caption[]{Properties of VLMSs and BDs during 
D-burning.\label{table1}  }
\begin{tabular}{rrrrrll}
\hline
\hline\noalign{\smallskip}
M  &  age & $D/D_0$ & $\te$ & $\log L$ &  $P_0$  & $\taud$  \\
$(\msol)$  & (Myr) & & (K) & ($L_\odot$) &   (hr)  & (Myr) \\
\hline\noalign{\smallskip}
 0.02 &  2.22 & 0.982 & 2540 & $-$2.516 & 1.67 & 3.03 \\
      & 15.9 & 0.054 & 2450  & $-$2.801 & 1.18 & 2.49 \\
      & 18.9 & 0.002 & 2390 & $-$2.927 & 1.04 & 25.1 \\
\hline\noalign{\smallskip}
0.03  & 1.18 & 0.936 &  2710 & $-$2.138 & 2.16 & 0.78 \\
      &  6.60 & 0.121  & 2690  & $-$2.271 & 1.77 & 0.94 \\
      & 8.49 & 0.004 & 2680 & $-$2.380 & 1.52 & 8.02 \\
\hline\noalign{\smallskip}
0.04  & 1.23 & 0.66 & 2790 & $-$1.915 & 2.52 & 0.41 \\
      & 3.71 & 0.148 & 2790   & $-$1.985 & 2.26 & 0.65 \\
      &  5.16 & 0.004 &  2780  & $-$2.092 & 1.92 & 5.92 \\
\hline\noalign{\smallskip}
0.06  & 0.86 & 0.644 & 2890 & $-$1.567 & 3.41 & 0.31 \\
      & 2.16 & 0.218 & 2890  & $-$1.616 & 3.15 & 0.43 \\
      & 3.36 & 0.005 & 2890 & $-$1.728 & 2.63 & 4.10 \\
\hline\noalign{\smallskip}
0.08  & 0.65 & 0.642 & 2950 & $-$1.322 & 4.25 & 0.26 \\
         & 1.39 & 0.315 & 2940 & $-$1.356 & 4.02 & 0.32 \\
         & 2.54 & 0.006 & 2940 & $-$1.475 & 3.27 & 3.32 \\
\hline\noalign{\smallskip}
0.1   & 0.56 & 0.618 & 3010  & $-$1.125 & 5.04 & 0.23 \\
      & 1.11 & 0.307 &  3000 & $-$1.162 & 4.77 & 0.29 \\
      & 2.00 & 0.008 & 3000  & $-$1.281 & 3.92 & 2.40 \\
\hline
\end{tabular}
\end{table*}

\end{document}